\documentclass[a4paper]{article}
\usepackage{graphics,graphicx,amsmath,amsfonts,lscape, url, mathtools, bm}
\usepackage[usenames]{color}
\usepackage{subfig}
\usepackage[countmax]{subfloat}
\usepackage{rotating}

\newcommand{\expit}{\mbox{expit}}

\numberwithin{equation}{section}
\usepackage{multirow}

\begin{document}

\title{Subgroup analysis of treatment effects for misclassified biomarkers with time-to-event data}

\author{Fang Wan, Andrew C. Titman, Thomas F. Jaki}

\date{}
\maketitle

\begin{abstract}

Analysing subgroups defined by biomarkers is of increasing importance in clinical research. In many situations the biomarker is subject to misclassification error, meaning the subgroups are identified with imperfect sensitivity and specificity. In these cases, {
it is improper to assume the Cox proportional hazards model for the subgroup specific treatment effects for time-to-event data with respect to the true subgroups, since} the survival distributions with respect to the diagnosed subgroups will not adhere to the proportional hazards assumption. This precludes the possibility of using simple adjustment procedures. Instead, we present a method based on formally modelling the data as a mixture of Cox models using an EM algorithm for estimation. An estimate of the overall population treatment effect is obtained through the interpretation of the hazard ratio as a concordance odds. Profile likelihood is used to construct individual and simultaneous confidence intervals of treatment effects. The resulting confidence intervals are shown to have close to nominal coverage for moderately large sample sizes in simulations and the method is illustrated on data from a renal-cell cancer trial.

\end{abstract}

\section{Introduction}
There is increasing acknowledgement of the existence of patient subgroups within clinical research. While some treatments work well for all patients with the same disease, it has been shown that some treatments are only effective for some subgroups of patients defined by a certain predictive biomarker \cite{ Sacks, Jackson, Jimeno, Peeters}. As a consequence, many clinical trials look to perform subgroup analysis to assess whether a treatment is beneficial for those patients that are biomarker positive or biomarker negative and many trial designs have been developed to account for these {
 subgroups}. Enrichment designs \cite{Freidlin, Magnusson} seek to identify the most promising (sub)group of patients during the study while other designs optimize the cost-efficiency of the trials via patients allocation with respect to their biomarker status (subgroup membership) \cite{Wason, Mehta} or use a biomarker-strategy design \cite{Kunz}. All of these clinical trials assume $100\%$ accuracy of the biomarker used in defining subgroups. However, it is seldom possible to measure a biomarker with perfect diagnostic accuracy meaning the observed subgroups will be subject to misclassification error. Without taking the sensitivity and specificity of the biomarker into consideration, the resulting conclusion may be inaccurate \cite{Liu, Wan}.   


Existing methods that account for the sensitivity and specificity (see \cite{Liu, Wan}) consider normal and binary endpoints only, while time-to-event data has not yet been considered.

\begin{figure}
    \centering
\includegraphics[width=0.5\textwidth]{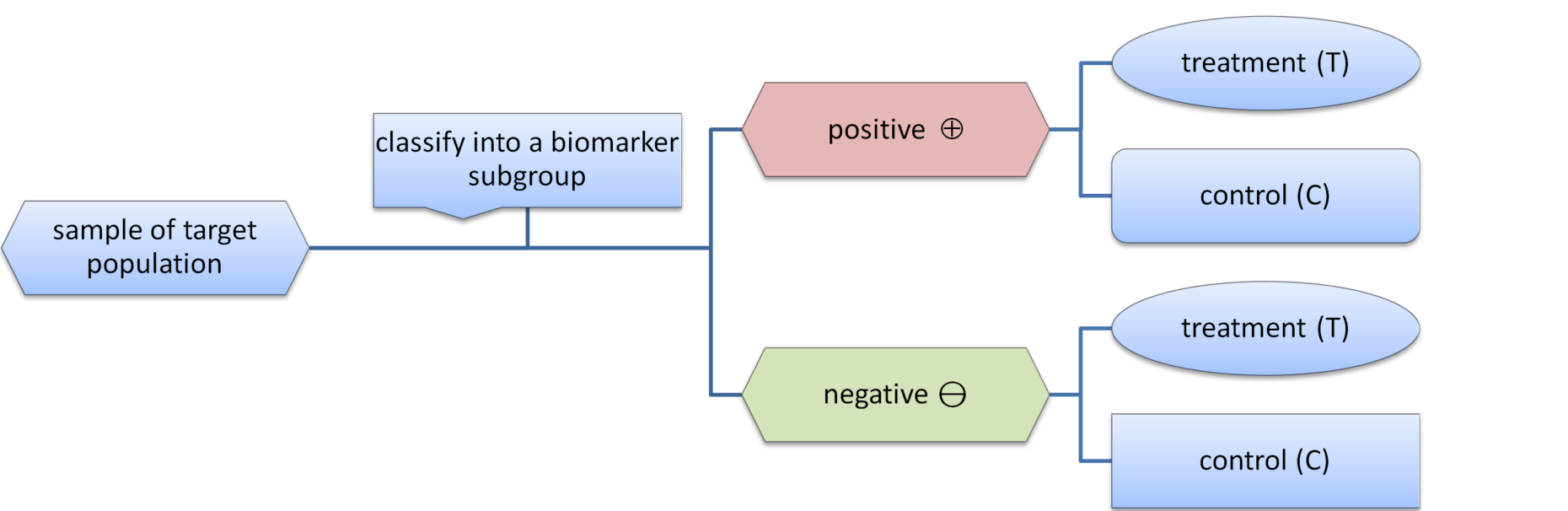}\\
\caption{A biomarker stratified design}
\label{graph0}
\end{figure}

In this paper, we propose a method to obtain point estimates and confidence intervals of the treatment effects in biomarker stratified subgroups with time-to-event data for a biomarker by treatment interaction design \cite{Gosho} (see Figure \ref{graph0}): Assume the total number of patients available to be enrolled into the trial is fixed to be $N$. Patients are classified into two subgroups according to the observed status (positive or negative) of a specific biomarker. In each of the two subgroups, patients are randomized into either the treatment or control arm and are administered experimental treatment or placebo/active control accordingly. The primary outcome, which is the survival time subject to right censoring, of all patients enrolled are recorded for analysis.

The remainder of the article is organized as follows. In Section 2, the statistical model for misclassified biomarker subgroups is defined. Section 3 gives estimation procedures for the model parameters, measures of overall efficacy and construction of confidence intervals. Section 4 presents a simulation results to assess the performance of the estimator and confidence intervals. The method is illustrated on a data example relating to metastatic renal-cell cancer in Section 5. The article concludes with a discussion.

\section{Statistical Model}
\label{statmod}

Conditional on the true biomarker status, a proportional hazards model is assumed to hold. Specifically the hazard at time $t$ for patient $i$ is taken as
\begin{equation}h_{i}(t; x_i, z_i) = h_{0}(t)\exp(\beta_1 x_i + \beta_2 z_i + \gamma x_i z_i)\label{model}\end{equation}
where $h_{0}(t)$ is an unspecified baseline hazard function, $x_i$ and $z_i$ are binary indicators of treatment and true biomarker status, respectively. Note that the biomarker status is 0 for the true negative subgroup and  1 for the true positive subgroup. {
Further note that this model assumes that the biomarker status to be measured without error in this model.} Under this model, the hazard ratios associated with the treatment are $\exp(\beta_1)$ and $\exp(\beta_1 + \gamma)$ for patients in the biomarker negative and positive group, respectively.

When the true biomarker status cannot be observed, a diagnostic test with imperfect sensitivity and specificity has to be used. Let $v_i \in \{0,1\}$ be a binary indicator of whether the $i$th patient tests positive for the biomarker. 
The {
marginal} distribution of survival times among patients in each diagnosis group will then be a mixture of Cox models corresponding to the models under true biomarker positive or negative status and with the mixing proportions determined by the positive-predictive value (PPV) and negative-predictive value (NPV) of the diagnostic test.

The PPV is given by
\begin{equation}\label{ppv}p_{+|\oplus}:= \frac{\pi \times \lambda_1}{\pi \times  \lambda_1 + (1- \pi)(1- \lambda_2)}\end{equation} and NPV by
\begin{equation}\label{npv}p_{-|\ominus}:=\frac{(1- \pi) \lambda_2}{\pi (1 -  \lambda_1) + (1- \pi)\lambda_2} \end{equation}
where the sensitivity, $\lambda_1$, and the specificity, $\lambda_2$ are assumed to be known and the prevalence of the biomarker, $\pi$, may either be considered known or will be estimated from the data. 

The survivor function for patients observed to be positive and negative are then 
\begin{equation}S_{\oplus}(t; x) \coloneqq S(t; x, v=1) = p_{+|\oplus} S(t ; x, z=1) + (1 - p_{+|\oplus}) S(t ; x, z=0)\label{splus}\end{equation} and
\begin{equation}S_{\ominus}(t; x) \coloneqq S(t; x, v=0) = (1 - p_{-|\ominus}) S(t ; x, z=1) + p_{-|\ominus} S(t ; x, z=0),\label{sminus}\end{equation} 
respectively, where
$S(t ; x, z) = \exp\{-H_{0}(t)\exp(\beta_1 x + \beta_2 z + \gamma x z)\}$ and $H_{0}(t) = \int_{0}^{t} h_{0}(u) du$ is the baseline cumulative hazard.

Note that unless there is either no treatment effect or the biomarker is observed without misclassification, proportional hazards will not hold with respect to the treatment $x$, for either $S_{\oplus}(t; x)$ or $S_{\ominus}(t;x)$. Therefore it is not possible to fit a Cox model to the observed data and perform some simple correction to adjust for misclassification error. Instead, a formal likelihood-based estimation procedure is used.

\section{Estimation}

Recently, Wu {\it et al} \cite{wu} proposed a Logistic-Cox mixture model in order to test for the existence of subgroups. In their model a logistic regression model determines the effect of observable covariates on the probability of membership of a mixture component. In each mixture component, times-to-event follow a Cox model with different covariate effects. 

The problem of correcting for misspecified biomarker status can be considered a special case of the framework of Wu {\it et al}, where the mixing probabilities have a specific form that is fully specified given the prevalence, $\pi$, and the sensitivity and specificity.

In order to estimate the model proposed in Section \ref{statmod}, a semi-parametric maximum likelihood approach is employed. A full likelihood for the data is constructed by making the standard assumption that the hazard function $h_{0}(t)$ is piecewise constant between observed event times \cite{breslow}. 

\subsection{EM algorithm}
\label{em_alg}

Direct maximization of the likelihood is difficult or infeasible due to the large number of nuisance parameters associated with the increments of the baseline hazard. Instead, taking a similar approach to various previous authors  \cite{zhong, martinussen}, an Expectation-Maximization (EM) algorithm is used. The true biomarker status is treated as missing data, such that the `M'-step of the algorithm involves fitting a weighted Cox model, where each patient has two sets of data corresponding to being truly biomarker positive or biomarker negative. The 
weights correspond to the conditional probability of being truly biomarker positive (or negative) given the current estimates of the parameters and the observed data (follow-up time, event indicator and diagnostic test result). 

Let $t_i$, for $i=1, \ldots, n$, denote the follow up time for patient $i$ and $\delta_i$ correspond to an event indicator, the likelihood contribution of the $i$th observed data given positive and negative subgroup status is
$$L_{+i} = [h_{0}(t_i) \exp\{(\beta_1 + \gamma)x_i + \beta_2\}]^{\delta_i} \exp[-H_{0}(t_i  ) \exp\{(\beta_1 + \gamma)x_i + \beta_2\}]$$
and
$$L_{-i} = [h_{0}(t_i) \exp\{\beta_1 x_i \}]^{\delta_i} \exp[-H_{0}(t_i) \exp\{\beta_1 x_i \}],$$
where $H_0(t)$ denotes the cumulative baseline hazard.

The conditional weights are then given by

\begin{eqnarray*}w_i &\coloneqq & P(z_i = 1 | t_i, \delta_i, \bm{\theta}, x_i, v_i, {H}_{0}(t))\\
&=& \left(\frac{p_{+|\oplus}L_{+i}}{ p_{+|\oplus}L_{+i} + (1 -p_{+|\oplus})L_{-i}} \right)^{v_i} \left(\frac{(1 - p_{-|\ominus})L_{+i}}{ (1 - p_{-|\ominus})L_{+i} + p_{-|\ominus}L_{-i}} \right)^{1 - v_i},\\
\end{eqnarray*}
where $\bm\theta = (\beta_1, \beta_2, \gamma)$.
Note that the weights depend on both $\bm\theta$ and $H_0$.  At each iteration, the M-step first updates the estimates of $\bm\theta$ and then updates $\hat{h}_{0}(t)$  and $\hat{H}_{0}(t)$ using the Breslow's estimator for the baseline hazard from the weighted Cox model \cite{breslow}. Let $t_{(j)}$ denote the $j$th ordered uncensored event time and $t_{(0)}=0$. Then
$$\hat{h}_0(t) = h_j, ~~\mbox{for}~~ t_{(j-1)} < t \leq t_{(j)},$$
where $$h_j = \left( \{t_{(j)} - t_{(j-1)}\} \sum_{l \in \mathcal{R}_j} w_l \exp\{(\beta_1 + \gamma)x_l + \beta_2\} + (1- w_l)\exp\{\beta_1 x_l \}\right)^{-1}$$
and $\mathcal{R}_j = \{i : t_i \geq t_{(j)} \}$ denotes the risk set of patients at time $t_{(j)}$.

These new estimates of $\bm\theta$ and $H_{0}(t)$ are subsequently used to update the weights. When the prevalence parameter $\pi$ is treated as unknown, it is also updated at each iteration, with the updated value of $\pi$ given by $n^{-1} \sum_{i} w_i$. This involves also updating the values of $p_{+|\oplus}$ and $p_{-|\ominus}$ by plugging the new estimate of $\pi$ into (\ref{ppv}) and (\ref{npv}).


The marginal, or observed, likelihood for the data is given by
\begin{equation*}
\label{prevunk}
L(\bm\theta, \hat{H}_0(t), \pi) =\prod_i\{P(t_i,\delta_i|x_i,\nu_i,\bm\theta, \hat{H}_0(t), \pi)P(\nu_i|\pi)\}.
\end{equation*}
If the prevalence of disease, $\pi$, is known then the above likelihood can be expressed as
\begin{equation}
\label{prevkn}
L(\bm\theta, \hat{H}_0(t)) = \prod_i \{p_{+|\oplus}L_{+i} + (1 - p_{+|\oplus})L_{-i}\}^{v_i} \{(1 -p_{-|\ominus})L_{+i} + p_{-|\ominus}L_{-i}\}^{1 - v_i},\end{equation}
which corresponds to the likelihood conditional on the observed diagnostic test results, $v_i$.
In the case where $\pi$ is treated as unknown, we have 
\begin{equation}
\label{prevunk}
\begin{split}
L(\bm\theta, \hat{H}_0(t), \pi) = \prod_i \{\pi \times \lambda_1 L_{+i} + (1-\pi)\times (1- \lambda_2)L_{-i}\}^{v_i} \times \\
\{\pi \times (1- \lambda_1) L_{+i} + (1-\pi) \times \lambda_2 L_{-i}\}^{1 - v_i},
\end{split}\end{equation}
with the difference between (\ref{prevkn}) and (\ref{prevunk}) arising because of the necessity to include terms relating to the probabilities of observed values of $v_i$ in the latter case. 
\subsection{Measures of overall efficacy}

The use of hazard ratios for subgroup analysis of time-to-event data has been criticized due to the absence of a constant hazard ratio in a mixture population and as a consequence, other methods based on median survival and parametric modelling have been proposed to obtain `subgroup mixable' estimates \cite{ding}.

However, the hazard ratio between two groups in a proportional hazards model can also be expressed as the concordance odds \cite{schemper}. Specifically, if $T_0$ and $T_1$ are the survival times of two randomly chosen individuals from groups 0 and 1 and the hazard ratio of group 1 compared to group 0 is $\psi$, then $\frac{P(T_0 > T_1)}{1 - P(T_0 > T_1)} = \psi$, or equivalently
\begin{equation}
P(T_0 > T_1) = \frac{\psi}{1 + \psi}.
\label{psi}
\end{equation} 
The concordance odds has a clear clinical meaning and has the advantage that an estimate of the overall concordance odds in a subgroup model can be found as a function of just the individual subgroup concordance odds and the prevalence. It also has the advantages of not requiring either fully parametric estimation procedures, which may be less robust, or fully non-parametric procedures which will be less efficient.

Let $T_i, i=0,1$ represent the survival time of a random subject in treatment arm $i$ and $G_i \in \{0,1\}, i=0,1$ represent the subgroup membership with $P(G_i=1) = \pi$, then
\begin{equation*}
\begin{split}
P(T_0 > T_1) = \pi^2 P(T_0 > T_1 | G_0=G_1=1) + (1-\pi)^2 P(T_0 > T_1 | G_0=G_1=0)~~~~~~\\~~~~~~ + \pi(1-\pi)P(T_0 > T_1 | G_0=0,G_1=1) + \pi(1-\pi)P(T_0 > T_1 | G_0=1,G_1=0),
\end{split}
\end{equation*}

hence
$$P(T_0 > T_1) = \pi^2 P(T_{01} > T_{11}) + (1-\pi)^2 P(T_{00} > T_{10}) + \pi(1-\pi)P(T_{00} > T_{11}) + \pi(1-\pi)P(T_{01} > T_{10})$$
where $T_{ij}$ is the survival time for a subject in treatment arm $i$ and subgroup $j$. 
Each of the probabilities on the right-hand side of the equation can be expressed in terms of the parameters $\beta_1, \beta_2, \gamma$ of the model in (\ref{model}). Following (\ref{psi}), we have
\begin{equation}P(T_0 > T_1) = \pi^2 \expit(\beta_1 + \gamma) + (1-\pi)^2 \expit(\beta_1) + \pi(1-\pi) \expit(\beta_1 + \beta_2 + \gamma) + \pi(1-\pi)\expit(\beta_1 - \beta_2),\label{conceq}\end{equation}
where $\expit(x) = (1 + \exp(-x))^{-1}$ is the inverse logit function.
An estimate of the overall effect of a treatment, expressed as concordance odds, is then given by $\frac{\hat{P}(T_0 > T_1)}{1 -\hat{P}(T_0 > T_1)}$ where $\hat{P}(T_0 > T_1)$ is obtained by plugging the estimates of $\beta_1, \beta_2, \gamma$ and $\pi$ into (\ref{conceq}).

A disadvantage of the concordance odds as a measure of treatment efficacy is that there is no guarantee that the overall efficacy measure lies in the interval between the two subgroup efficacy values. In fact, when $\gamma = 0$ but $\beta_2 \neq 0$ the overall concordance odds will be closer to 1 than $\exp(\beta_1)$. Nevertheless it is virtually impossible, in practice, for a contradictory result to occur, i.e. for there to be statistically significant benefits for both subgroups but a non-significant overall effect. Such a situation could only occur if $|\hat{\beta}_2|$ is large and $SE(\hat{\beta}_2)$ is large compared to both $SE(\hat{\beta}_1)$ and $SE(\hat{\beta}_1 + \hat{\gamma})$. Moreover it is impossible for the overall effect to be of a different sign to the two subgroup effects. 

\subsection{Construction of confidence intervals}
A convenient approach to constructing asymptotic confidence intervals for individual parameters is based upon the profile likelihood ratio, which continues to have standard $\chi^2$ asymptotics even in the presence of a potentially infinitely dimensional nuisance parameter \cite{murphy_annals}. For instance, to obtain a confidence interval for the interaction, $\gamma$, we use the fact that $$\Lambda(\hat\gamma, \gamma_0) = 2 \log\frac{L(\hat{\bm\theta}, \hat{H}_{0}(t))}{L(\hat{\beta}_1,\hat{\beta}_2,\gamma_0,
\hat{H}_{0}(t))} \xrightarrow{d} \chi^2_1$$ and hence take $$\{ \gamma : \Lambda(\hat\gamma, \gamma) \leq \chi^{2}_{1}(1 - \alpha) \}$$ as a $(1-\alpha) \times 100\%$ confidence interval for $\gamma$. It is straightforward to find the maximum profile likelihood estimates by using a modified EM algorithm where at each M-step the fixed parameter, e.g. $\gamma$, is treated as a fixed offset term in the weighted Cox model.

For the subgroup analysis it is also desirable to construct a simultaneous confidence interval for the estimated treatment effect in the biomarker positive and negative groups in order to control the familywise type I error rate. In the parametrization used in (\ref{model}) this corresponds to simultaneous confidence intervals for $(\beta_1 + \gamma)$ and $\beta_1$. In this case, the method proceeds by obtaining an estimate of the Hessian of the observed profile likelihood with respect to $(\beta_1, \gamma)$. The  method of Murphy and van der Vaart \cite{murphy} is used to approximate the profile likelihood information. This approach has also been used in other contexts where estimation requires an EM algorithm \cite{xu}.
The profile likelihood information is approximated by computing the profile likelihood at values about $(\hat\beta_1, \hat\gamma)$, perturbed by a suitably small value $h$ to provide a `finite-differences' type approximation.
Specifically,
$$I_{\beta_1\beta_1} \approx -\frac{l_{p}(\hat\beta_1 + 2h, \hat\gamma) - 2l_{p}(\hat\beta_1 + h, \hat\gamma) + l_{p}(\hat\beta_1, \hat\gamma)}{h^2},$$
$$I_{\beta_1\gamma} \approx -\frac{l_{p}(\hat\beta_1 + h, \hat\gamma + h) - l_{p}(\hat\beta_1, \hat\gamma+ h)-l_{p}(\hat\beta_1+h, \hat\gamma) + l_{p}(\hat\beta_1, \hat\gamma)}{h^2}$$ and
$$I_{\gamma\gamma} \approx -\frac{l_{p}(\hat\beta_1 , \hat\gamma+2h) - 2l_{p}(\hat\beta_1, \hat\gamma+h) + l_{p}(\hat\beta_1, \hat\gamma)}{h^2}.$$
The value of $h$ is primarily chosen to ensure that numerical stability in the converged values of the EM algorithm do not affect the estimate. Theoretically, the value of $h$ should decrease with increasing sample size, but taking $h=0.01$ worked adequately in the examples considered in this paper and the results were not particularly sensitive to the choice of $h$.

By inverting the estimated information matrix $\mathbf{I}$,
an estimate of the variance-covariance matrix of $(\hat{\beta}_1 + \hat\gamma,\hat{\beta}_1)$,  $$\Sigma = \bigl( \begin{smallmatrix} \sigma^2_{+}& \rho \sigma_{+}\sigma_{-}\\
\rho \sigma_{+}\sigma_{-} & \sigma^2_{-} \end{smallmatrix} \bigr),$$
 is given by using the delta method
 $$\hat\Sigma = \bigl( \begin{smallmatrix}1& 1\\1& 0  \end{smallmatrix} \bigr)\mathbf{I}^{-1}\bigl( \begin{smallmatrix}1 &1\\1& 0  \end{smallmatrix} \bigr).$$ Simultaneous confidence intervals are then constructed of the form
$$(\hat{\beta}_1 + \hat\gamma) \pm \xi_{\alpha} \sigma_{+}$$ and
$$\hat{\beta}_1 \pm \xi_{\alpha} \sigma_{-}$$ where  $\xi_{\alpha}$ is the scaling factor chosen such that, for a bivariate normal random variable, $\mathbf{X}$, with unit variances and correlation $\rho$, $P(|X_1| \leq  \xi_{\alpha} \cap |X_2| \leq  \xi_{\alpha}) = 1 - \alpha$. This value can be found straightforwardly using the \verb!qmvnorm! function in the mvtnorm package in {\bf R} \cite{mvtnorm, genz}.
The procedure can be extended to provide simultaneous confidence intervals for the two subgroup effects and the overall log concordance odds by computing the variance-covariance matrix of $(\hat{\beta}_1 + \hat{\gamma},\hat{\beta}_1, \hat{\beta}^{*})$ using the delta method, where $\hat{\beta}^{*} = \log\left\{\frac{\hat{P}(T_0 > T_1)}{1 -\hat{P}(T_0 > T_1)}\right\}$. Obtaining an analytical form for the first derivatives can be cumbersome, but a numerical approximation for the first derivatives can be used instead.

\subsection{Missing biomarker status}

In some trials, only a subset of patients may have had their biomarker status measured. 
If it can be assumed that the missing diagnostic tests of biomarker status are missing at random, then the survivor function for such patients, $S_{\odot}(t;x)$, is given by:
$$S_{\odot}(t;x) = \pi S(t;x,z=1) + (1- \pi)S(t;x,z=0).$$
Such patients can easily be accommodated within the EM algorithm proposed in Section \ref{em_alg} by a simple modification of the conditional weights for such patients. Specifically, the weight for a patient $i$ with missing diagnostic test is taken as
$$w_i = \frac{\pi L_{+i}}{\pi L_{+i} + (1 - \pi)L_{-i}}.$$  
Similarly, the marginal likelihood contribution of these patients, regardless of whether $\pi$ is taken as known or to be estimated is simply given by
$$L_i(\bm\theta, \hat{H}_0(t), \pi) = \pi L_{+i} + (1- \pi)L_{-i}.$$

\section{Simulations}

To investigate the finite sample properties of the proposed estimator data sets of varying sizes and levels of biomarker subgroup diagnostic accuracy are simulated. The underlying survival hazards are assumed to follow the model in (\ref{model}), with a decreasing Weibull baseline hazard assumed such that $h_0(t) = 0.8 \times 0.1^{0.8} t^{-0.2}$.

Three scenarios are considered for the treatment effects. In the first, $\beta_1 = -0.5, \beta_2 = 0.1$ and $\gamma = 0.3$, meaning the treatment is beneficial for both biomarker groups, but the effect is smaller for those who are biomarker positive, corresponding to hazard ratios (HR) of 0.61 and 0.82.
In the second, $\beta_1 = 0.1, \beta_2 = 0.1$ and $\gamma = -0.7$ corresponding to a stronger interaction effect where the treatment is beneficial for the biomarker positive group but slightly harmful for the negative group (HRs of 0.55 and 1.11).
Finally the third scenario, $\beta_1 = 0, \beta_2 = 0.1$ and $\gamma = 0$, corresponds to a situation where the treatment has no effect in either biomarker group.

Censoring is assumed to be independent and uniform distributed between 5 and 25, $\mbox{U}(5,25)$, which results in an overall censoring rate of around 25\%.
The prevalence of a true positive biomarker status, $\pi$, is taken to be 0.3 and treated as unknown in the estimation procedure. The effect of assuming rather than estimating the prevalence is negligible in the simulation (where the prevalence given is accurate). However, if the prevalence given is far from its true value, there should be an impact. The sensitivity and specificity of the diagnostic test and the sample size per randomization group is varied across the simulation scenarios. 
The results from 5000 replications of each scenario are presented in Tables \ref{tab_sim1}, \ref{tab_sim2} and \ref{tab_sim3}.
The parameter estimates have reasonably low levels of bias for all scenarios considered. As would be expected, the standard deviation of the estimates increases as the diagnostic accuracy decreases. In the scenarios considered, since the prevalence is lower than 0.5, imperfect specificity has a greater impact than imperfect sensitivity. The standard deviation of the estimate of $\gamma$ is around 75\% higher when the sensitivity and specificity are both 0.8 compared to the case of perfect diagnostic accuracy.  In the first scenario where the interaction effect is relatively modest, the power to detect the interaction term is low in all scenarios, but substantially lower when there is diagnostic error. For instance when the number in each randomization group is 500, the power reduces from 0.43, with perfect diagnostic accuracy, to 0.17 with sensitivity and specificity both 0.8. A similar pattern is observed in the second scenario, where the interaction effect is stronger. In the scenario with no interaction effect the empirical Type I error of a test of interaction is close to 5\% for all configurations, with a slight tendency to be anti-conservative in the smaller sample size and when misclassification rates are higher.

\begin{table}[ht]
\caption{Bias and Standard Deviation (SD) of parameter estimates, empirical coverage of simultaneous (Simult) nominal 95\% confidence intervals of $(\beta_1, \beta_1 + \gamma)$ and the empirical power of likelihood ratio test of interaction term in the mild interaction scenario $(\beta_1, \beta_2, \gamma) = (-0.5, 0.1, 0.3)$.}
\label{tab_sim1}
\begin{center}
\begin{tabular}{cc|ccc|ccc|cc}
&&\multicolumn{3}{c|}{Bias $\times 10^2$}&\multicolumn{3}{c|}{SD}&Coverage&Power\\
$N$&(Sens, Spec)&$\beta_1$&$\beta_2$&$\gamma$&$\beta_1$&$\beta_2$&$\gamma$&Simult&$\gamma \neq 0$\\
\hline
100&(1,1)&0.0191&0.7232&-0.6650&0.2153&0.2634&0.3847&0.9472&0.1226\\
100&(1,0.8)&-0.4598&0.9980&-2.0488&0.2416&0.3514&0.5154&0.9614&0.0902\\
100&(0.8,1)&-0.6302&1.1681&-1.1618&0.2299&0.3017&0.4473&0.9512&0.1128\\
100&(0.9,0.9)&-0.6628&1.1261&-1.2436&0.2383&0.3366&0.4960&0.9622&0.0912\\
100&(0.8,0.8)&-0.4748&0.0643&-1.7780&0.2685&0.4767&0.6887&0.9596&0.0814\\
500&(1,1)&0.1274&0.2206&-0.2761&0.0965&0.1157&0.1670&0.9506&0.4286\\
500&(1,0.8)&-0.1734&-0.1977&-0.0315&0.1077&0.1583&0.2280&0.9546&0.2490\\
500&(0.8,1)&-0.3087&-0.1782&0.4508&0.1012&0.1333&0.1948&0.9522&0.3514\\
500&(0.9,0.9)&-0.2142&-0.3322&0.1361&0.1052&0.1490&0.2172&0.9520&0.2834\\
500&(0.8,0.8)&-0.1233&-0.4452&-0.3492&0.1218&0.2051&0.2949&0.9578&0.1678\\
\end{tabular}
\end{center}
\end{table}

\begin{table}[ht]
\caption{Bias and Standard Deviation (SD) of parameter estimates, empirical coverage of simultaneous (Simult) nominal 95\% confidence intervals of $(\beta_1, \beta_1 + \gamma)$ and the empirical power of likelihood ratio test of interaction term in the strong interaction scenario $(\beta_1, \beta_2, \gamma) = (0.1, 0.1, -0.7)$.}
\label{tab_sim2}
\begin{center}
\begin{tabular}{cc|ccc|ccc|cc}
&&\multicolumn{3}{c|}{Bias $\times 10^2$}&\multicolumn{3}{c|}{SD}&Coverage&Power\\
$N$&(Sens, Spec)&$\beta_1$&$\beta_2$&$\gamma$&$\beta_1$&$\beta_2$&$\gamma$&Simult&$\gamma \neq 0$\\
\hline
100&(1,1)&0.1509&0.1085&-0.5872&0.2034&0.2581&0.3966&0.9468&0.4562\\
100&(1,0.8)&-0.6065&-0.5878&1.0292&0.2210&0.3430&0.5144&0.9570&0.2968\\
100&(0.8,1)&0.0270&1.2441&-1.6333&0.2132&0.3004&0.4587&0.9516&0.3682\\
100&(0.9,0.9)&-0.4779&-0.8028&-0.1945&0.2241&0.3650&0.5453&0.9578&0.2812\\
100&(0.8,0.8)&-1.5560&0.6694&-1.3342&0.253&0.4714&0.8217&0.9502&0.2004\\
500&(1,1)&0.1167&0.1671&-0.5071&0.0900&0.1148&0.1705&0.9522&0.9878\\
500&(1,0.8)&0.1796&0.3993&-0.3036&0.0967&0.1480&0.2173&0.9566&0.9042\\
500&(0.8,1)&0.1194&0.1749&-0.1344&0.0926&0.1306&0.1975&0.9536&0.9452\\
500&(0.9,0.9)&0.1950&0.4669&-0.4286&0.0981&0.1563&0.2282&0.9556&0.8772\\
500&(0.8,0.8)&-0.0108&0.6640&-0.0728&0.1126&0.2010&0.2959&0.9600&0.6752\\
\end{tabular}
\end{center}
\end{table}

\begin{table}[ht]
\caption{Bias and Standard Deviation (SD) of parameter estimates, empirical coverage of simultaneous (Simult) nominal 95\% confidence intervals of $(\beta_1, \beta_1 + \gamma)$ and the empirical Type I error of likelihood ratio test of interaction term in the null scenario $(\beta_1, \beta_2, \gamma) = (0, 0.1, 0)$.}
\label{tab_sim3}
\begin{center}
\begin{tabular}{cc|ccc|ccc|cc}
&&\multicolumn{3}{c|}{Bias $\times 10^2$}&\multicolumn{3}{c|}{SD}&Coverage&Type I err\\
$N$&(Sens, Spec)&$\beta_1$&$\beta_2$&$\gamma$&$\beta_1$&$\beta_2$&$\gamma$&Simult&$\gamma \neq 0$\\
\hline
100&(1,1)&0.1389&0.0680&0.5389&0.2012&0.2629&0.3678&0.9508&0.0496\\
100&(1,0.8)&-0.1232&0.5505&0.1944&0.222&0.3356&0.4798&0.9580&0.0588\\
100&(0.8,1)&-0.0146&0.2379&1.2609&0.2155&0.3109&0.4396&0.9492&0.0620\\
100&(0.9,0.9)&-0.0436&-0.8576&1.3769&0.2257&0.3644&0.5110&0.9590&0.0588\\
100&(0.8,0.8)&-0.3072&0.6792&0.1749&0.2541&0.4698&0.7010&0.9608&0.0662\\
500&(1,1)&0.1800&0.4851&-0.4689&0.0899&0.1171&0.1657&0.9436&0.0572\\
500&(1,0.8)&0.2906&0.3950&-0.4087&0.0971&0.1480&0.2067&0.9528&0.0482\\
500&(0.8,1)&0.1894&0.1761&-0.0657&0.0928&0.1307&0.1848&0.9526&0.0508\\
500&(0.9,0.9)&0.3246&0.4625&-0.5343&0.0994&0.1563&0.2196&0.9556&0.0472\\
500&(0.8,0.8)&0.3481&0.6671&-0.630&0.1128&0.2011&0.2832&0.9622&0.0476\\
\end{tabular}
\end{center}
\end{table}

The simultaneous confidence intervals for $(\beta_1, \beta_1 + \gamma)$ have close to the nominal 95\% level in all cases, with a tendency to be slightly conservative.

\section{Example: Pazonpanib for renal-cell cancer}

As an illustrative example of the impact of accounting for misclassification of biomarkers in a survival study, data from a Phase III trial of patients with metastatic renal-cell cancer are analyzed. The trial involved 343 patients, 225 of whom were randomized to treatment with Pazopanib, with the remaining 118 on placebo. In addition, patients were classified by level of interleukin 6 (IL-6) into `low' or `high' groups. Interest lies in determining whether Pazonpanib is an effective treatment for either or both groups of patient. In the original analysis by Tran {\it et al} \cite{tran}, it was assumed that the assay used to determine the level of IL-6 had 100\% diagnostic sensitivity and specificity. 

Here, the data are re-analysed considering the possibility of misclassification of IL-6 status. The individual level data were reconstructed from the Kaplan-Meier estimates provided in Tran {\it et al} \cite{tran} using the method of Guyot {\it et al} (2012) \cite{Guyot}.
Following Liu {\it et al} \cite{Liu}, it is assumed that the assay has 95\% sensitivity and 90\% specificity to distinguish high IL-6 from low. 

Table \ref{tab1} compares the results of an analysis assuming no misclassification with estimates using the proposed method. It is seen that the effect of adjusting for misclassification is to increase the estimated interaction effect from -0.53 to -0.72, which also leads to the interaction being considered significant ($p=0.036$).

\begin{table}[ht]
\caption{Comparison of estimates from original Cox model analysis assuming no biomarker misclassification and model assuming 95\% sensitivity and 90\% specificity to detect High IL-6}
\label{tab1}
\begin{center}
\begin{tabular}{c|ccc|ccc}
& \multicolumn{3}{c|}{Original analysis}&\multicolumn{3}{c}{Misclassification corrected}\\
Parameter & Estimate & 95\% CI & $p$-value & Estimate  & 95\% CI & $p$-value \\
\hline
Pazonpanib ($\beta_1$) &-0.15& (-0.58, 0.27) & 0.48&  -0.12 &(-0.57, 0.33) & 0.58\\
High IL-6 ($\beta_2$) & 1.18 & (0.73, 1.62)& $<0.001$& 1.50 &(0.96, 2.10)& $<0.001$ \\
Interaction ($\gamma$) & -0.53 & (-1.08, 0.03)& 0.06& -0.72 & (-1.40, -0.05)& 0.04\\
Prevalence ($\pi$) & - & - & - & 0.47 & (0.41, 0.53)& -\\
\end{tabular}
\end{center}
\end{table}

Table \ref{tab2} gives the estimates and simultaneous 95\% confidence intervals for the concordance odds of Pazonpanib for Low and High IL-6 patients. For both the original and misclassification analyses, the confidence interval for Low IL-6 includes 1, implying no treatment effect, whilst the confidence interval for High IL-6 is entirely below 1, indicating a treatment benefit. 
%
%

\begin{table}[ht]
\caption{Simultaneous 95\% confidence intervals for effect of Pazonpanib on overall survival for Low IL-6 patients, High IL-6 patients and all patients; CO=Concordance odds}
\label{tab2}
\begin{center}
\begin{tabular}{c|cc|cc}
& \multicolumn{2}{c|}{Original analysis}&\multicolumn{2}{c}{Misclassification corrected}\\
Group & CO & 95\% CI & CO & 95\% CI \\
\hline
Low IL-6 & 0.86 & (0.52, 1.41)& 0.88 & (0.52, 1.50)\\
High IL-6 & 0.51 & (0.33, 0.77)& 0.43 & (0.25, 0.75) \\
All &  0.70  & (0.51 , 0.95) & 0.67 & (0.50 , 0.92) \\
\end{tabular}
\end{center}
\end{table}

\section{Discussion}
In this paper, we investigate subgroup analysis for time-to-event responses in biomarker stratified subgroups with misclassificated biomarkers using a proportional hazards model. Point estimation and the construction of (simultaneous) confidence intervals for the treatment effects in biomarker subgroups in the form of the log-hazard ratio are provided. It is shown by simulation that the bias of the estimators and the coverage probabilities of the simultaneous confidence intervals are acceptable for all considered simulation scenarios. 

It is also apparent from the simulation results that the power to detect a subgroup effect of treatment is diminished in the presence of misclassification. Further work would be to develop sample size formulas which would allow survival trials to be adequately powered to perform subgroup analysis in the presence of biomarker misclassification.

The interpretation of a hazard ratio as the concordance odds allows an overall treatment effect estimate to be computed in subgroup analyses of time-to-event data. While the focus of this paper has been cases with misclassification of the biomarker status, the use of concordance odds can also be applied in the simpler case where the biomarker status is perfectly observed. 


\section{Acknowledgement}
This work is an independent research arising in part from Prof Jaki's Senior Research Fellowship (NIHR-SRF-2015-08-001) supported by the National Institute for Health Research. Funding for this work was also provided by the Medical Research Council (MR/M005755/1). The views expressed in this publication are those of the authors and not necessarily those of the NHS, the National Institute for Health Research, or the Department of Health.


\begin{thebibliography}{unsrt}
\bibitem{Sacks} Sacks FM, Pfeffer MA, Moye LA, Rouleau JL, Rutherford JD, Cole TG, Braunwald E. The effect of Pravastatin on coronary events after myocardial infarction in patients with average cholesterol levels. {\it New England Journal of Medicine.} 1996. {\bf 335}:1001-1009. DOI:10.1056/NEJM199610033351401.

\bibitem{Jackson} Jackson RD, LaCroix AZ, Gass M, Wallace RB, Robbins J, Lewis CE, Barad D. Calcium plus vitamin D supplementation and the risk of fractures. {\it New England Journal of Medicine.} 2006. {\bf 354}:669-683. DOI:10.1056/NEJMoa055218.


\bibitem{Jimeno} Jimeno A, Messersmith WA, Hirsch FR, Franklin WA, Eckhardt SG. KRAS mutatioons and sensitivity to epidermal growth factor receptor inhibitors in colorectal cancer: practical application of patient selection. {\it Journal of Clinical Oncology}.2009.{\bf 27}(7): 1130-1136. DOI:


\bibitem{Peeters} Peeters M, Douillard JY, Van Cutsem E, Siena S, Zhang K, Williams R, Wiezorek J. Mutant KRAS codon 12 and 13 alleles in patients with metastatic colorectal cancer: assessment as prognostic and predictive biomarkers of response to panitumumab. {\it Journal of Clinical Oncology}.2013.{\bf 31}(6): 759-765. DOI:

\bibitem{Freidlin} Freidlin B, Korn EL. Biomarker enrichment strategies: matching trial design to biomarker credentials.{\it Nat Rev Clin Oncol}. 2014. {\bf 11}(2):81-90. DOI: 10.1038/nrclinonc.2013.218. 

\bibitem{Magnusson} Magnusson BP, Turnbull BW. Group sequential enrichment design incorporating subgroup selection. {\it Statistics in Medicine}. 2013. {\bf 32}(16): 2695–2714. DOI: 10.1002/sim.5738.

\bibitem{Wason} Wason JMS, Abraham JE, Baird RD, Gournaris I, Vallier AL, Brenton JD, Earl HM, Mander AP. A Bayesian adaptive design for biomarker trials with linked treatments. {\it British Journal of Cancer}. 2015. 113, 699-705. DOI: 10.1038/bjc.2015.278.

\bibitem{Mehta} Mehta C, Schafer H, Daniel H, Irle S. Biomarker driven population enrichment for adaptive oncology trials with time to event endpoints. {\it Statistics in Medicine.} 2014. {\bf 33}(26): 4515–4531. DOI: 10.1002/sim.6272.

\bibitem{Kunz} Kunz C, Jaki T, Stallard N. An alternative method to analyse the biomarker-strategy design.  Submitted.

\bibitem{Liu} Liu, C., Liu, A., Hu, J., Yuan, V. and Halabi, S. Adjusting for misclassification in a stratified biomarker clinical trial. {\it Statistics in Medicine} 2014. {\bf 33}: 3100-3113.

\bibitem{Wan} Wan F, Kunz C, Jaki T. Confidence regions for treatment effects in subgroups in biomarker stratified designs. Submitted.

\bibitem{Gosho} Gosho M, Nagashima K, Sato Y. Study designs and statistical analyses for biomarker research. {\it Sensors.} 2012. 12: 8966-8986. DOI: 10.3390/s120708966.

\bibitem{wu} Wu RF, Zheng M, Yu W. Subgroup Analysis with Time-to-Event Data Under a Logistic-Cox Mixture Model. {\it Scandinavian Journal of Statistics}. 2016. DOI: 10.1111/sjos12213.

\bibitem{breslow} Breslow NE, Discussion of the paper by D.R. Cox. {\it Journal of the Royal Statistical Society; Series B}. 1972. {\bf 34}: 216-217.

\bibitem{zhong} Zhong M, Sen PK, Cai J. (1996). Cox regression model with mismeasured covariates or missingcovariate data. {\it ASA Biometrics Section Proceedings.} 323-328.

\bibitem{martinussen} Martinussen T. Cox regression with Incomplete Covariare Measurements using the EM- algorithm. {\it Scandinavian Journal of Statistics}. 1999. {\bf 26}: 479-491.

\bibitem{Bilmes} Bilmes JA. A Gentle Tutorial of the EM Algorithm and its Application to Parameter Estimation for Gaussian Mixture and Hidden Markov Models. {\it  International Computer Science Institute; Technical Report.} 1998. TR-97-021.
\bibitem{ding} Ding Y, Lin H-M, Hsu JC. Subgroup mixable inference on treatment efficacy in mixture populations, with an application to time-to-event outcomes. {\it Statistics in Medicine}. 2015. {\bf 35}: 1580-1594. DOI: 10.1002/sim.6822.

\bibitem{schemper} Schemper M, Wakounig S, Heinze G. The estimation of average hazard ratios by weighted Cox regression. {\it Statistics in Medicine}. 2009; {\bf 28}: 2473–2489. DOI: 10.1002/sim.3623.

\bibitem{murphy_annals} Murphy SA, van der Vaart AW. Semiparametric likelihood ratio inference. {\it Annals of Statistics}.1997. {\bf 25}: 
1471-1509.

\bibitem{murphy} Murphy SA, van der Vaart AW. Observed information in semiparametric models. {\it Bernoulli}. 1999. {\bf 5}: 381-412.

\bibitem{xu} Xu C, Baines PD, Wang, JL. Standard error estimation using the EM algorithm for the joint modeling of survival and longitudinal data. {\it Biostatistics}. 2014. {\bf 15}: 731-744.

\bibitem{mvtnorm} Genz A, Bretz F, Miwa T, Mi X, Leisch F, Scheipl F, Hothorn T. mvtnorm: Multivariate Normal and t Distributions. {\it R package version 0.9-9997.} 2014 \url{http://CRAN.R-project.org/package=mvtnorm}

\bibitem{genz} Genz A, Bretz F. Computation of Multivariate Normal and t Probabilities. {\it Lecture Notes in Statistics}. 2009. {\bf 195}. Springer-Verlage, Heidelberg.



\bibitem{tran} Tran HT, Liu Y, Zurita AJ, Lin Y, Baker-Neblett KL, Martin AM, Figlin RA, Hutson TE, Sternberg CN, Amado RG, Pandite LN. Prognostic or predictive plasma cytokines and angiogenic factors for patients treated with pazopanib for metastatic renal-cell cancer: a retrospective analysis of phase 2 and phase 3 trials. {\it The Lancet Oncology} 2012. {\bf 13}: 827-837.

\bibitem{Guyot} Guyot P, Ades A, Ouwens MJ, Welton NJ. Enhanced secondary analysis of survival data: reconstructing the data from published Kaplan-Meier survival
curves. {\it BMC Medical Research Methodology}. 2012. {\bf 12}:9.


\end{thebibliography}
\end{document}